\documentclass[12pt]{article}%
\usepackage{amsmath}
\usepackage{amsfonts}
\usepackage{amssymb}
\usepackage{graphicx}%
\providecommand{\U}[1]{\protect\rule{.1in}{.1in}}
\begin{document}

\title{A Compact Treatment of the Friedel-Anderson and the Kondo Impurity Using the
FAIR Method \\(\textbf{F}riedel \textbf{A}rtificially \textbf{I}nserted \textbf{R}esonance)}
\author{Gerd Bergmann\\Department of Physics\\University of Southern California\\Los Angeles, California 90089-0484\\e-mail: bergmann@usc.edu}
\date{\today}
\maketitle

\begin{abstract}
Although the Kondo effect and the Kondo ground state of a magnetic impurity
have been investigated for more than forty years it was until recently
difficult if not impossible to calculate spatial properties of the ground
state. In particular the calculation of the spatial distribution of the
so-called Kondo cloud or even its existence have been elusive. In recent years
a new method has been introduced to investigate the properties of magnetic
impurities, the FAIR method, where the abbreviation stands for Friedel
Artificially Inserted Resonance. The FAIR solution of the Friedel-Anderson and
the Kondo impurity problems consists of only eight or four Slater states.
Because of its compactness the spatial electron density and polarization can
be easily calculated. In this article a short review of the method is given. A
comparison with results from the large N-approximation, the Numerical
Renormalization Group theory and other methods shows excellent agreement. The
FAIR solution yields (for the first time) the electronic polarization in the
Kondo cloud.
\[
\]
\newpage

\end{abstract}

\section{Introduction}

The properties of magnetic impurities in a metal is one of the most
intensively studied problems in solid state physics. The work of Friedel
\cite{F28} and Anderson \cite{A31} laid the foundation to understand why some
transition-metal impurities form a local magnetic moment while others don't.
Kondo \cite{K8} showed that multiple scattering of conduction electrons by a
magnetic impurity yields a divergent contribution to the resistance in
perturbation theory. Kondo's paper stimulated a large body of theoretical and
experimental work which changed our understanding of d- and f-impurities
completely (see for example \cite{Y2}, \cite{V7}, \cite{S77}, \cite{D44},
\cite{H23}, \cite{M20}, \cite{A36}, \cite{G24}, \cite{C8}, \cite{H20}). A
large number of sophisticated methods were applied in the following three
decades to better understand and solve the Kondo and Friedel-Anderson
problems. In particular, it was shown that at zero temperature the
Friedel-Anderson impurity is in a non-magnetic state. To name a few of these
methods: scaling \cite{A51}, renormalization \cite{W18}, \cite{F30},
\cite{K58}, \cite{K59} Fermi-liquid theory \cite{N14}, \cite{N5}, slave-bosons
(see for example \cite{N7}), large-spin limit \cite{G19}, \cite{B103}. After
decades of research exact solutions of the Kondo and Friedel-Anderson
impurities were derived with help of the Bethe-ansatz \cite{W12}, \cite{A50},
\cite{S29}, representing a magnificent theoretical achievement. The
experimental and theoretical progress has been collected in a large number of
review articles \cite{D44}, \cite{H23}, \cite{M20}, \cite{A36}, \cite{G24},
\cite{C8}, \cite{H20}, \cite{W18}, \cite{N5}, \cite{N7}, \cite{B103},
\cite{W12}, \cite{A50}, \cite{S29}, \cite{N17}.

The exact theory of the Bethe ansatz is such a complex theory that only a
limited number of parameters can be calculated. For the majority of practical
problems one uses the numerical renormalization group (\textbf{NRG}) theory
and the large-spin (large $N$) method. Recently the author introduced another
approximate solution for the Friedel-Anderson (FA) \cite{B152}, \cite{B151}
and the Kondo impurity \cite{B153}, the FAIR method. The FAIR solution
consists of only four to eight Slater states and is therefore very compact. It
is well suited to calculate in particular spatial properties of the Kondo
ground state. It yields the first quantitative calculations of the Kondo cloud
\cite{B177}. There are very few spatial properties of the Kondo ground state
calculated with other theoretical approaches. One example is the NRG
calculation for the Friedel oscillations in the vicinity of a Kondo impurity
\cite{A83}. A calculation of the Friedel oscillations with the FAIR method
yields good agreement with the NRG results \cite{B178}. In this short review
the FAIR method will be introduced and some of the results presented. The FAIR
method uses Wilson states \cite{W18} which replace a complete conduction
electron band by a relatively small number of states which carry the full
interaction with the impurity. The Wilson states are sketched in appendix A.

\section{II The FAIR Method\smallskip}

\subsection{The artificial Friedel resonance state}

We consider the Hamiltonian of a band with a finite number $N$ of
non-interacting electron states
\[
H_{0}=%
{\textstyle\sum_{\nu=1}^{N}}
\varepsilon_{\nu}c_{\nu}^{\dag}c_{\nu}%
\]
The $c_{\nu}^{\dag}$ are the creation operators of the band. \emph{In the
following the states such as }$c_{\nu}^{\dag}\Phi_{0}$ \emph{are represented
and addressed by their creation operators }$c_{\nu}^{\dag}$\emph{\ suppressing
the vacuum states }$\Phi_{0}$\emph{. }From these band states a new (arbitrary)
state $a_{0}^{\dag}$ is composed
\begin{equation}
a_{0}^{\dag}=\sum_{\nu=1}^{N}\alpha_{0}^{\nu}c_{\nu}^{\dag} \label{a0}%
\end{equation}
In the next step an intermediate basis $\left\{  \overline{a}_{i}^{\dag
}\right\}  $ can be constructed numerically where the additional $\left(
N-1\right)  $ states $\overline{a}_{i}^{\dag}$ are orthonormal to each other
and to $a_{0}^{\dag}$. In this basis the Hamiltonian $H_{0}$ is given by an
$N\times N$ matrix with the elements $\left(  H_{0}\right)  _{ij}$ where
$\left(  H_{0}\right)  _{00}$ is at the left upper corner. In the final step
the $\left(  N-1\right)  \times\left(  N-1\right)  $ sub-matrix of $\left(
H_{0}\right)  _{ij}$ for $i,j>0$ is diagonalized. This yields the new basis
$\left\{  a_{i}^{\dag}\right\}  =\left\{  a_{0}^{\dag},a_{1}^{\dag}%
,a_{2}^{\dag},..,a_{N-1}^{\dag}\right\}  $ which is uniquely determined by the
state $a_{0}^{\dag}$. In this basis the s-band Hamiltonian has the form
\begin{equation}
H_{0}=\sum_{\nu=1}^{N}E_{i}a_{i}^{\dag}a_{i}+E_{0}a_{0}^{\dag}a_{0}+\sum
_{\nu=1}^{N}V_{fr}\left(  i\right)  \left(  a_{0}^{\dag}a_{i}+a_{i}^{\dag
}a_{0}\right)  \label{H0}%
\end{equation}
One recognizes that this Hamiltonian is analogous to a Friedel Hamiltonian
where $a_{0}^{\dag}$ represents an artificial Friedel resonance. Therefore
this state is called a FAIR state for \textbf{F}riedel \textbf{A}rtificially
\textbf{I}nserted \textbf{R}esonance state.

It has to be emphasized that the FAIR state $a_{0}^{\dag}$ can have any
composition of the basis states $c_{\nu}^{\dag}$. Therefore it can be adjusted
to a given problem without any restriction. This gives the FAIR method its adaptability.

\subsection{The Friedel resonance}

As an example let us consider the simple Friedel resonance Hamiltonian
$H_{Fr}$.
\begin{equation}
H_{FR}=\sum_{\nu=1}^{N}\varepsilon_{\nu}c_{\nu}^{\dag}c_{\nu}+E_{d}d^{\dag
}d+\sum_{\nu=1}^{N}V_{sd}(\nu)[d^{\dag}c_{\nu}+c_{\nu}^{\dag}d] \label{H_Fr}%
\end{equation}
Since $H_{FR}$ does not depend on the spin the latter will be ignored. The
first term is the conduction band Hamiltonian $H_{0},$ the second term gives
the energy of the d (resonance) state of the impurity with $d^{\dag}$ being
its creation operator. The last term represents the interaction between the d
state and the conduction electrons.

There exists a FAIR state $a_{0}^{\dag}$ and a FAIR basis $\left\{
a_{i}^{\dag}\right\}  $ so that the $n$-electron ground state of the Friedel
Hamiltonian is exactly given by%

\begin{equation}
\Psi_{Fr}=\left(  Aa_{0}^{\dag}+Bd^{\dag}\right)
{\textstyle\prod\limits_{i=1}^{n-1}}
a_{i}^{\dag}\Phi_{0} \label{Psi_F}%
\end{equation}
Here $A$ and $B$ are coefficients which fulfill the condition $A^{2}+B^{2}=1$.
Actually this exact form of the ground state of the Friedel impurity can be
understood without any analytic or numerical calculation \cite{B92}. This is
shown in the appendix D. In ref. \cite{B91} it was discovered by a variation
of $a_{0}^{\dag}$ minimizing the ground state energy of the state
(\ref{Psi_F}) with respect to the Friedel Hamiltonian (\ref{H_Fr}). The state
$a_{0}^{\dag}$ determines all the other basis states $a_{i}^{\dag}$ uniquely.
Since the new basis $\left\{  a_{i}^{\dag}\right\}  $ has the same number of
states as the original basis $\left\{  c_{\nu}^{\dag}\right\}  $ the
construction of the basis $\left\{  a_{i}^{\dag}\right\}  $ is only possible
if the number $N$ of basis states is small. For $N\thickapprox10^{23}$ it
would be hard to construct the orthonormal sub-diagonal basis $\left\{
a_{i}^{\dag}\right\}  .$ Wilson has shown in his Kondo paper \cite{W18} how to
construct a finite basis $\left\{  c_{\nu}^{\dag}\right\}  $ which preserves
the full interaction with the impurity. The Wilson states are discussed in the
appendix A.

The example of the Friedel Hamiltonian shows the simplicity and effectiveness
of the FAIR method. It can be applied to treat the Friedel-Anderson and the
Kondo impurity.\medskip

\section{The Friedel-Anderson impurity}

The FA-Hamiltonian consists of the Friedel Hamiltonian (\ref{H_Fr}) for both
spins plus a Coulomb term of the form $H_{\text{C}}=Un_{d\uparrow
}n_{d\downarrow}$.%
\begin{equation}
H_{FA}=%
{\textstyle\sum_{\sigma}}
\left\{  \sum_{\nu=1}^{N}\varepsilon_{\nu}c_{\nu\sigma}^{\dag}c_{\nu\sigma
}+E_{d}d_{\sigma}^{\dag}d_{\sigma}+\sum_{\nu=1}^{N}V_{sd}(\nu)[d_{\sigma
}^{\dag}c_{\nu\sigma}+c_{\nu\sigma}^{\dag}d_{\sigma}]\right\}  +Un_{d\uparrow
}n_{d\downarrow} \label{H_FA}%
\end{equation}

\subsection{The magnetic state}

In the early years (before the Kondo paper) it was the goal to calculate (and
measure) the magnetic moment of a d- or f-impurity. After the discovery of the
Kondo effect and after Schrieffer and Wolff \cite{S31} transformed the
FA-Hamiltonian into a Kondo Hamiltonian it became clear that the ground state
of the FA impurity is non-magnetic. Then the calculation of the magnetic
moment was often considered as irrelevant, even heresy. The paper by
Krishna-murthy, Wilkins, and Wilson \cite{K58} clarified the role of the local
magnetic moment in the FA-impurity. KWW performed a numerical renormalization
a la Wilson \cite{W18} for the FA-Hamiltonian. They demonstrated that the
renomalization-group flow diagram showed very different flows from the
free-orbital fixed point H$_{\text{FO}}^{\ast}$ to the strong coupling fixed
point H$_{\text{SC}}^{\ast}$ (see Fig.1). For sufficiently large Coulomb
repulsion (when $U>>\Gamma=\pi\rho\left\vert V_{sd}\right\vert ^{2}$) the flow
of their Hamiltonian $H_{N}$ passed close to the (unstable) fixed point
H$_{\text{LM}}^{\ast}$ for a local moment. This means that under these
conditions the impurity assumed first a magnetic moment when the temperature
is lowered. After passing the fixed point for the local moment H$_{\text{LM}%
}^{\ast}$ the renormalization flow is essentially the same as for a Kondo
Hamiltonian (where a local moment is the starting point).%

\[%
\begin{array}
[c]{c}%
{\includegraphics[
height=2.5637in,
width=2.7364in
]%
{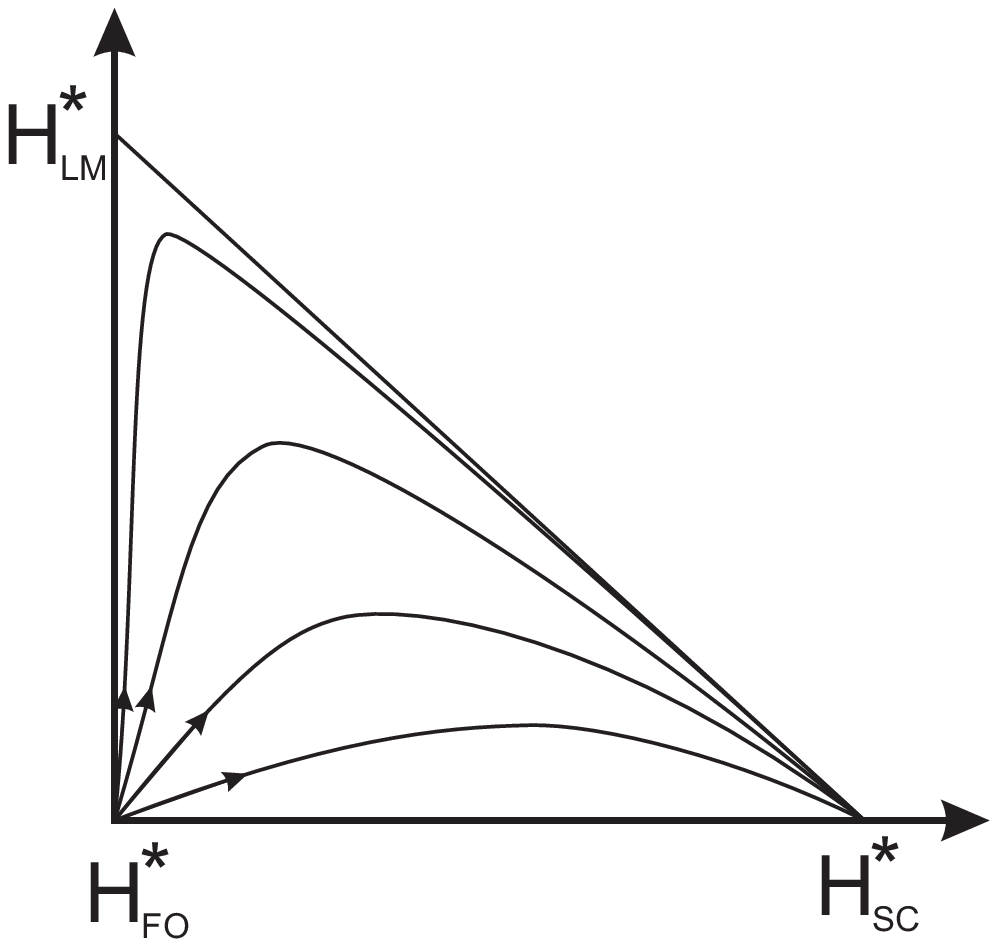}%
}%
\end{array}%
\begin{tabular}
[c]{l}%
Fig.1:Schematic\\
renormalization-group flow\\
diagram after ref.\cite{K58}.
\end{tabular}
\ \
\]%
\[
\]

With decreasing ratio of $U/\Gamma$ the flow path passes less and less close
to H$_{\text{LM}}^{\ast}$. This means that the size of the local moment
decreases until there is no longer a local moment formed. (The flow of the
susceptibility indicates this behavior).

At the end point of the renormalization the system approaches the strong
coupling fixed point H$_{\text{SC}}^{\ast}$ and\ shows the universal behavior
of the Kondo ground state. Nevertheless the ground-state wave functions are
quite different for small and large ratios of $U/\Gamma$ because the size of
the magnetic moment is engraved into the wave function.

Let us first consider the local moment state of the FA-impurity. This state is
a ground state if one applies a magnetic field which is strong enough to
suppress the Kondo ground state. Within the FAIR approach the (potentially)
magnetic solution has the form%
\begin{equation}
\Psi_{MS}=\left[  A_{a,b}a_{0\uparrow}^{\dag}b_{0\downarrow}^{\dag}%
+A_{a,d}a_{0\uparrow}^{\dag}d_{\downarrow}^{\dag}+A_{d,b}d_{\uparrow}^{\dag
}b_{0\downarrow}^{\dag}+A_{d,d}d_{\uparrow}^{\dag}d_{\downarrow}^{\dag
}\right]  \prod_{i=1}^{n-1}a_{i\uparrow}^{\dag}\prod_{i=1}^{n-1}%
b_{i\downarrow}^{\dag}\Phi_{0} \label{y1}%
\end{equation}
where $\left\{  a_{i}^{\dag}\right\}  $ and $\left\{  b_{i}^{\dag}\right\}  $
are two (different) FAIR bases of the $N$-dimensional Hilbert space. The state
(\ref{y1}) opens a wide playing field for optimizing the solution: (i) The
FAIR states $a_{0\uparrow}^{\dag}$ and $b_{0\downarrow}^{\dag}$ can be
individually optimized, each one defining a whole basis $\left\{  a_{i}^{\dag
}\right\}  ,$ $\left\{  b_{i}^{\dag}\right\}  $ and (ii) the coefficients
$A_{a,b},A_{a,d},A_{d,b},A_{d,d}$ can be optimized fulfilling only the
normalization condition $A_{a,b}^{2}+A_{a,d}^{2}+A_{d,b}^{2}+A_{d,d}^{2}=1$.
Since the relative size of the coefficients $A_{a,b},A_{a,d},A_{d,b}$ and
$A_{d,d}$ is not restricted this solution describes correlation effects well.
The optimization procedure is described in detail in the appendix B.

Fig.2 shows the structure of the four Slater states of $\Psi_{MS}$
graphically. The FAIR states $a_{0}^{\dag}$ and $b_{0}^{\dag}$ are imbedded in
the spin-up and down bands while the $d_{\uparrow}^{\dag}$ and $d_{\downarrow
}^{\dag}$ states are shown on the left and right side of the bands. Full
circles represent occupied states.%

\begin{align*}
&
\raisebox{-0pt}{\includegraphics[
height=1.3856in,
width=5.056in
]%
{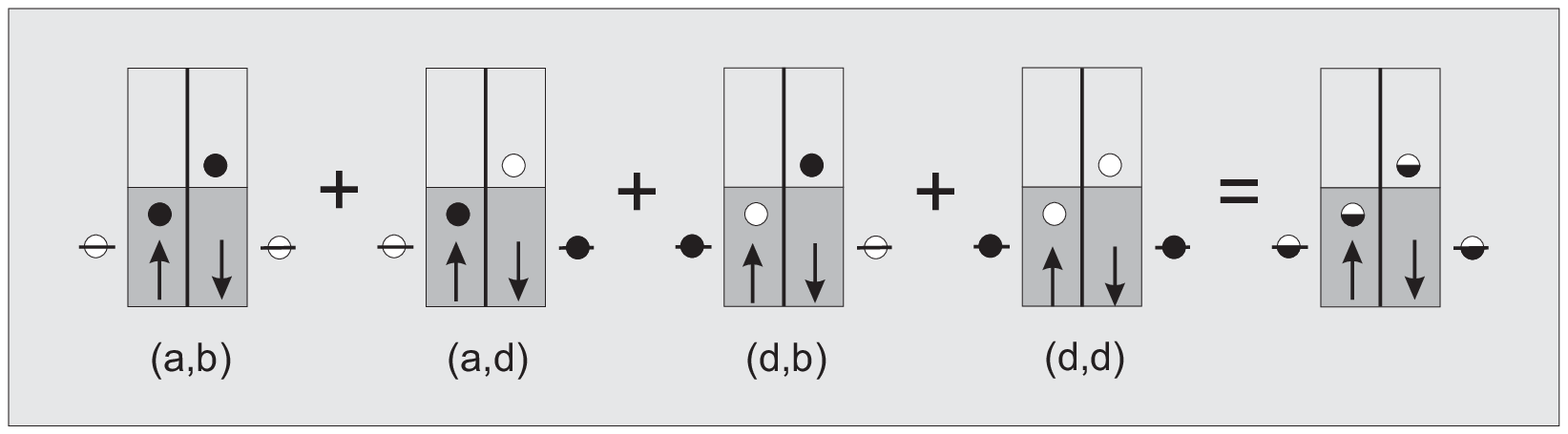}%
}%
\\
&
\begin{tabular}
[c]{l}%
Fig.2: The composition of the magnetic state $\Psi_{MS}$ is shown. It
consists\\
of four Slater states. Each Slater state has a half-filled spin-up and\\
down band, two FAIR states (circles in within the bands) and two d-states\\
(circles on the left and right of the band). Full black circles represent\\
occupied states and white circles represent empty states. The band\ at\\
the right with the half-filled circles symbolizes the magnetic solution with\\
four Slater states.
\end{tabular}
\end{align*}

Fig.3a shows the magnetic moment as a function of $U$ for the mean-field
solution and the magnetic state $\Psi_{MS}$. In Fig.3b the ground-state
energies of the mean-field solution and the magnetic state are compared. The
magnetic FAIR solution has a considerably lower energy expectation value than
the mean-field solution. More importantly it increases the critical value of
$U$ for the formation of a magnetic moment by almost a factor two (compared
with the mean-field solution). The mean-field approximation is still used in
combination with spin-density functional theory (SDFT) to calculate the
magnetic moment of impurities \cite{K46}, \cite{K47}, \cite{M41}, \cite{D28},
\cite{D33}. A combination between SDFT and the FAIR solution appears to be
very desirable.%

\begin{align*}
&
\begin{array}
[c]{cc}%
{\includegraphics[
height=2.1685in,
width=2.5645in
]%
{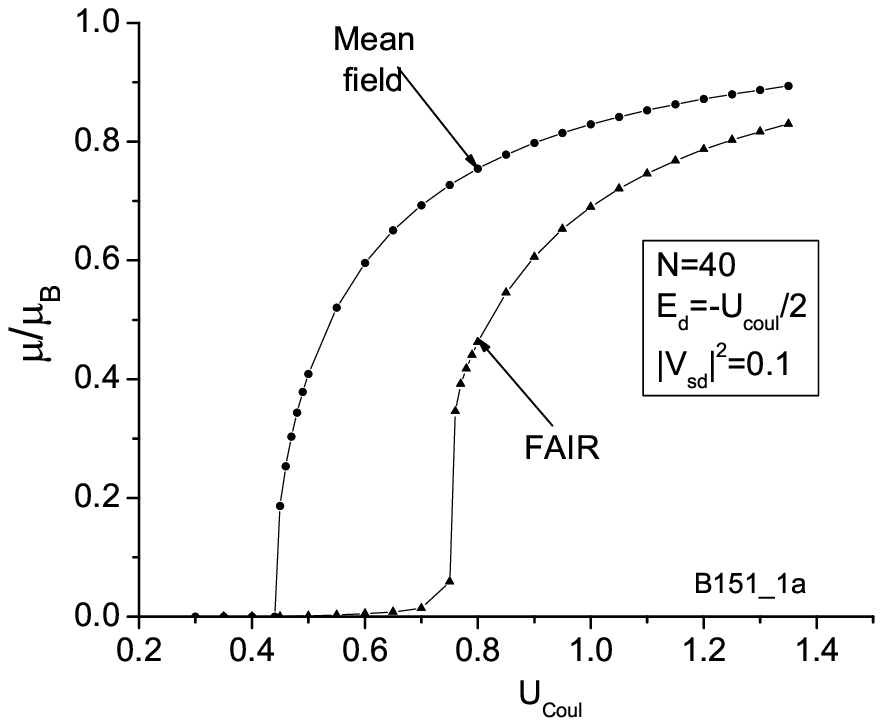}%
}%
&
{\includegraphics[
height=2.1403in,
width=2.5637in
]%
{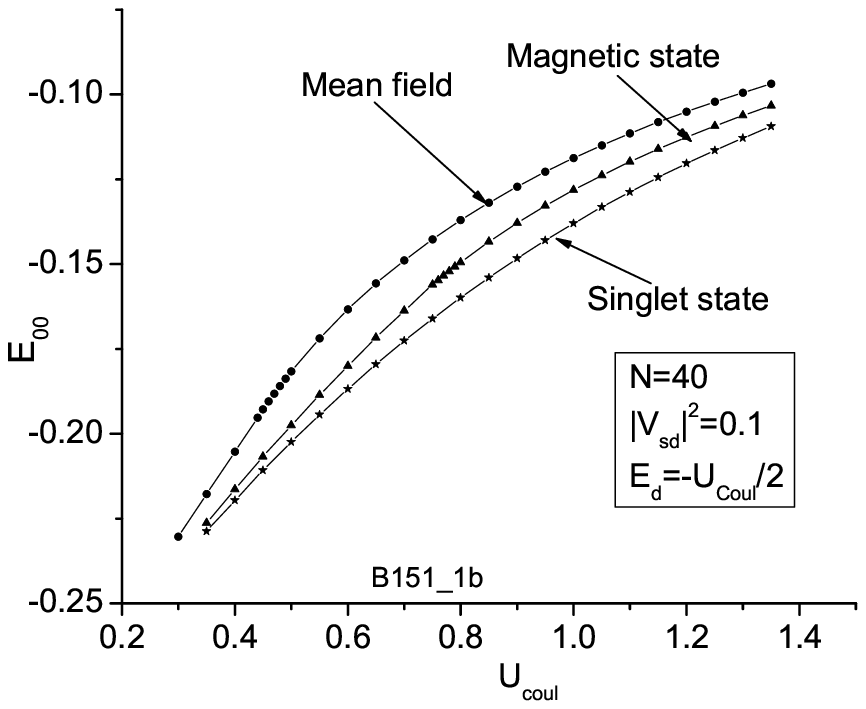}%
}%
\end{array}
\\
&
\begin{tabular}
[c]{l}%
Fig.3a: The magnetic moment as a function of the Coulomb exchange energy
$U,$\\
using the mean-field solution and the magnetic FAIR solution $\Psi_{MS}.$\\
Fig.3b: The ground-state energies of the mean-field solution, the magnetic
and\\
the singlet FAIR solution
\end{tabular}
\end{align*}

\subsection{The singlet state}

In hindsight it is quite natural that the magnetic state with its broken
symmetry is not the ground state. By reversing all spins one obtains a new
state with the same energy. In Fig.4 the energy of the magnetic state is
plotted as a function of the magnetic moment.
\begin{align*}
&
{\includegraphics[
height=2.225in,
width=2.6874in
]%
{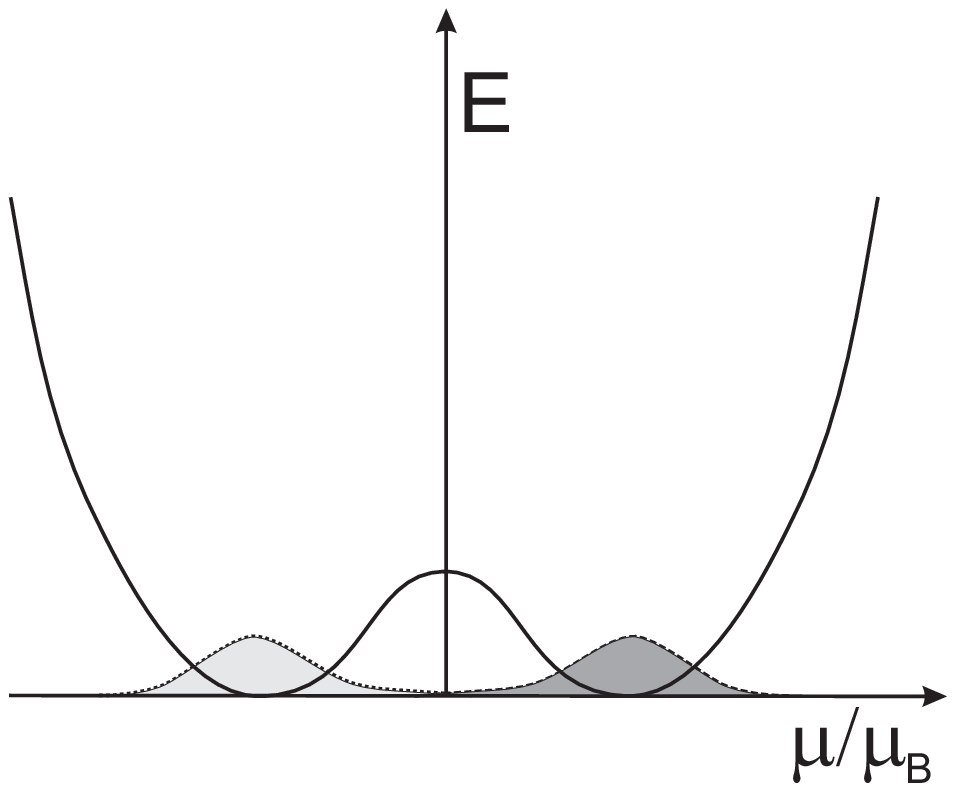}%
}%
\\
&
\begin{tabular}
[c]{l}%
Fig.4: The energy of the FA impurity as a function\\
of the magnetic moment.
\end{tabular}
\end{align*}

This is a situation similar to an atom in a double well potential. In the
ground state the atom is in a symmetric superposition of the wave functions in
the two wells. In analogy one can construct the singlet ground state
$\Psi_{SS}$ of the FA-Hamiltonian. This state is obtained by reversing all
spins in (\ref{y1}) and combining the two wave functions%
\[
\Psi_{SS}=\overline{\Psi_{MS}\left(  \uparrow\right)  }+\overline
{\overline{\Psi_{MS}\left(  \downarrow\right)  }}%
\]%
\begin{align}
&  =\left[  \overline{A_{a,b}}a_{0\uparrow}^{\dag}b_{0\downarrow}^{\dag
}+\overline{A_{a,d}}a_{0\uparrow}^{\dag}d_{\downarrow}^{\dag}+A_{d,b}%
d_{\uparrow}^{\dag}b_{0\downarrow}^{\dag}+A_{d,d}d_{\uparrow}^{\dag
}d_{\downarrow}^{\dag}\right]  \prod_{i=1}^{n-1}a_{i\uparrow}^{\dag}%
\prod_{i=1}^{n-1}b_{i\downarrow}^{\dag}\Phi_{0}\label{PsiSS}\\
&  +\left[  \overline{\overline{A_{a,b}}}b_{0\uparrow}^{\dag}a_{0\downarrow
}^{\dag}+\overline{\overline{A_{a,d}}}d_{\uparrow}^{\dag}a_{0\downarrow}%
^{\dag}+\overline{\overline{A_{d,b}}}b_{0\uparrow}^{\dag}d_{\downarrow}^{\dag
}+\overline{\overline{A_{d,d}}}d_{\downarrow}^{\dag}d_{\uparrow}^{\dag
}\right]  \prod_{i=1}^{n-1}b_{i\uparrow}^{\dag}\prod_{i=1}^{n-1}%
a_{i\downarrow}^{\dag}\Phi_{0}\nonumber
\end{align}

Again one has to optimize \thinspace$a_{0}^{\dag},$ $b_{0}^{\dag}$ and all the
coefficients. It is remarkable that the composition of the FAIR states changes
dramatically for small energies. The ground-state energy of the singlet FAIR
solution lies considerably below that of the magnetic FAIR solution $\Psi
_{MS}$ (see Fig.3b). One can compare this ground-state energy with a set of
numerical calculations by Gunnarsson and Schoenhammer \cite{G34}. They applied
the large $N_{f}$ method to the (non-degenerate) FA-Hamiltonian (spin $1/2$)
for a finite Coulomb interaction and included double occupancy of the impurity
level. They calculated the ground-state energy in the $1/N_{f}$-expansion up
to the order $\left(  1/N_{f}\right)  ^{2}$ which includes more than $10^{7}$
basis states. For the s-d-hopping transition they used an elliptic form. With
the following parameters: band width $B_{GS}=6eV$, Coulomb energy $U_{GS}%
=5eV$, d-state energy $E_{d,GS}=-2.5eV$ they performed two calculations, one
for s-d coupling $V_{GS}=1eV$ and another for $V_{GS}=2eV.$ The table compares
the ground-state energies and the occupation for of the d-states ($d_{0}%
,d_{1},d_{2}$) obtained by GS and the FAIR for $V_{GS}=1eV$ and $2eV$. Not
only the ground-state energies but also the occupation of the d-states agree
remarkably well.%

\begin{align*}
&
\begin{array}
[c]{cc}%
V_{GS}=1eV &
\begin{tabular}
[c]{|l|l|l|l|l|l|}\hline
\textbf{states} & $E_{0}\left[  \text{eV}\right]  $ & $d_{0}$ & $d_{1}$ &
$d_{2}$ & no. of coeff.\\\hline
GS & -0.245 & 0.034 & 0.931 & 0.034 &
$>$%
$10^{7}$\\\hline
FAIR & -0.239 & 0.035 & 0.931 & 0.034 & 80\\\hline
\end{tabular}
\\
& \\
V_{GS}=2eV &
\begin{tabular}
[c]{|l|l|l|l|l|l|}\hline
\textbf{states} & $E_{0}\left[  eV\right]  $ & $d_{0}$ & $d_{1}$ & $d_{2}$ &
no. of coeff.\\\hline
GS & -1.217 & 0.137 & 0.732 & 0.132 & $>10^{7}$\\\hline
FAIR & -1.234 & 0.140 & 0.722 & 0.138 & 80\\\hline
\end{tabular}
\end{array}
\\
&
\begin{tabular}
[c]{l}%
Table Ia,b: The ground-state energy $E_{0}$ and the occupations $d_{0}%
,d_{1},d_{2}$\\
of the d-states with 0,1 or 2 electrons.
\end{tabular}
\end{align*}

It is worthwhile to remember that the FAIR solution is completely determined
by the two FAIR states $a_{0}^{\dag}$ and $b_{0}^{\dag}$, i.e. by $2\times40$
amplitudes for a typical value of $N=40$. On the other hand the large N
calculation describes the ground state by more than $10^{7}$ parameter, i.e.,
amplitudes of Slater states.

\section{The Kondo Impurity{\protect\Large \smallskip}}

The Kondo Hamiltonian is a limiting case of the FA-Hamiltonian \cite{S31}. It
applies when the exchange energy $U$ approaches infinity while the d-state
energy approaches $-\infty$, for example $E_{d}=-U/2$. Then the d-state is
always singly occupied, either with spin up or down. The interaction between
the spin $\mathbf{s}$ of a conduction electron and the spin $\mathbf{S}$ of
the impurity can be expressed in the form $2J\mathbf{s\cdot S}$ where $J>0$.

In this case the ansatz for the compact FAIR-solution can be obtained from
equ. (\ref{PsiSS}). The coefficients $\overline{A_{a,b}},\overline
{\overline{A_{a,b}}}$; $A_{d,d},\overline{\overline{A_{d,d}}}\ $have to vanish
because there is only single occupancy of the d-state in the Kondo solution.
This yields%
\begin{align}
\Psi_{K}  &  =\left[  \overline{A_{a,d}}a_{0\uparrow}^{\dag}d_{\downarrow
}^{\dag}+\overline{A_{d,b}}d_{\uparrow}^{\dag}b_{0\downarrow}^{\dag}\right]
\prod_{i=1}^{n-1}a_{i\uparrow}^{\dag}\prod_{i=1}^{n-1}b_{i\downarrow}^{\dag
}\Phi_{0}\label{Psi_K}\\
&  +\left[  \overline{\overline{A_{a,d}}}d_{\uparrow}^{\dag}a_{0\downarrow
}^{\dag}+\overline{\overline{A_{d,b}}}b_{0\uparrow}^{\dag}d_{\downarrow}%
^{\dag}\right]  \prod_{i=1}^{n-1}b_{i\uparrow}^{\dag}\prod_{i=1}%
^{n-1}a_{i\downarrow}^{\dag}\Phi_{0}\nonumber
\end{align}

Again one can optimize the localized states $a_{0}^{\dag}$ and $b_{0}^{\dag}$
and the coefficients $\overline{A_{a,d}},\overline{\overline{A_{a,d}}}$,
$\overline{A_{d,b}},\overline{\overline{A_{d,b}}}$. If one arranges the spin
up to the left and spin down to the right in all components of (\ref{PsiSS})
then one obtains in the ground state $\overline{A_{a,d}}=\overline
{\overline{A_{a,d}}}$ and $\overline{A_{d,b}}=\overline{\overline{A_{d,b}}}.$
Our group calculated the total spin of this state (for $J=0.1$) and obtained
for the expectation value of $\left\langle \mathbf{S}^{2}\right\rangle
=\left\langle \left(  \mathbf{s}_{d}+%
{\textstyle\sum_{i}}
\mathbf{s}_{i}\right)  ^{2}\right\rangle $ the value $0.04$ in the ground
state \cite{B153}. For the first excited state one obtains $\left\langle
\mathbf{S}^{2}\right\rangle =1.99$. This means that the ground state is
essentially a singlet state ($\left\langle \mathbf{S}^{2}\right\rangle =0$)
and the first excited state a triplet state ($\left\langle \mathbf{S}%
^{2}\right\rangle =2$).

In the Kondo effect one is generally not so much interested in the
ground-state energy but in the so-called Kondo energy. This is, for example,
the energy difference between the triplet and singlet states. The logarithm of
this excitation energy is plotted in Fig.5 as a function of $1/\left(
2J\rho_{0}\right)  $ as the full circles ($\rho_{0}$ is the density of
states). The straight line corresponds to $\Delta E\thickapprox$
$5D\exp\left[  -1/\left(  2J\rho_{0}\right)  \right]  $. This is the unrelaxed
singlet-triplet energy which uses for the triplet state the same bases
$\left\{  a_{i}^{\dag}\right\}  $ and $\left\{  b_{i}^{\dag}\right\}  $ as in
the singlet state. One can derive a relaxed triplet state by the following
trick. In the triplet state the coefficients are related by $\overline
{\overline{A_{a,d}}}=-\overline{A_{a,d}}$ and $\overline{\overline{A_{d,b}}%
}=-\overline{A_{d,b}}$. If one replaces $\overline{\overline{A_{a,d}}},$
$\overline{\overline{A_{d,b}}}$ from the start by $-\overline{A_{a,d}%
},-\overline{A_{d,b}}$ and optimizes the energy then one obtains the relaxed
triplet energy. The difference between this energy and the singlet energy
yields the relaxed excitation energy $\Delta E_{st}$. (Since these are two
independent calculations they have to be performed with an absolute accuracy
of $10^{-10}$). This relaxed excitation energy is plotted in Fig.5 as stars.
The stars lie between two theoretical curves: (i) $\Delta E_{st}=D\exp\left[
-1/\left(  2J\rho_{0}\right)  \right]  ,$ given by the dashed curve and (ii)
$\Delta E_{st}=$ $\sqrt{2J\rho_{0}}D\exp\left[  -1/\left(  2J\rho_{0}\right)
\right]  $, given by the dotted curve. Both expressions are given in the
literature as approximate values for the Kondo temperature $k_{B}T_{K}$. The
numerical values lie closer to the second expression. Therefore the relaxed
singlet-triplet excitation energy corresponds closely to the Kondo energy and
confirms that the FAIR method represents the physics of the Kondo impurity accurately.%

\begin{align*}
&
\begin{array}
[c]{c}%
{\includegraphics[
height=3.868in,
width=4.7131in
]%
{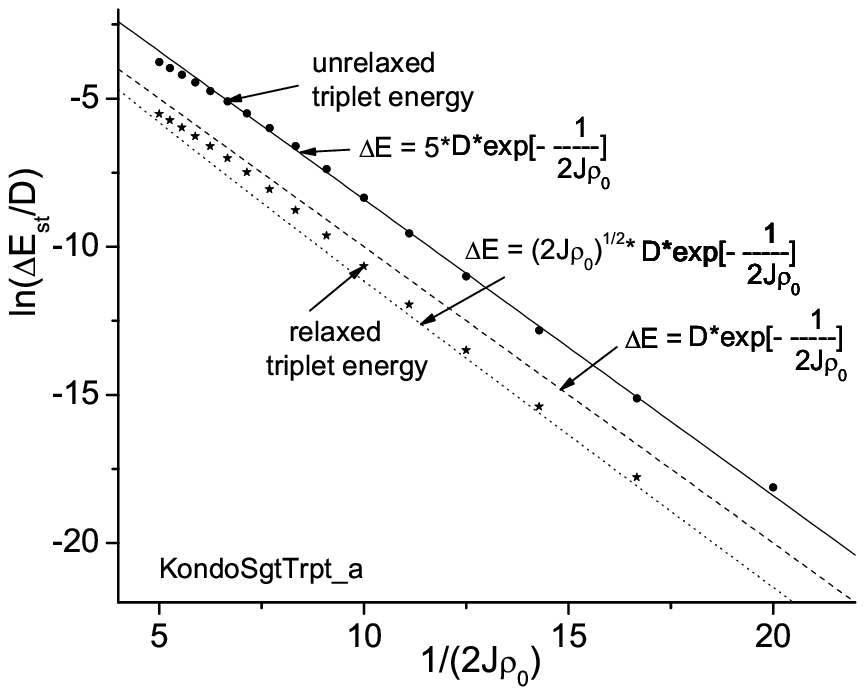}%
}%
\end{array}
\\
&
\begin{tabular}
[c]{l}%
Fig.5: The energy difference between the singlet and triplet\\
states. The full circles represent the unrelaxed singlet-triplet\\
excitation energy $\Delta E_{st}$ while the stars yield the relaxed
singlet-triplet\\
excitation energy $\Delta E_{st}^{\ast}$ (see text). The dashed and dotted
curves are\\
theoretical expressions for the Kondo energy.
\end{tabular}
\ \ \ \ \ \
\end{align*}

The FAIR method yields \textbf{both} energies, the ground-state and the
singlet-triplet excitation energy, with good accuracy although the two
energies can differ by a factor of thousand.

\section{Real Space Properties}

Since the compositions of the magnetic state and the singlet state are
explicitly known and consist only of a few Slater states it is straight
forward to calculated the electron density and spin polarization of the
different states. The details of the calculation are described in ref.
\cite{B177}.

In Wilson's approach the wave number $k$ is given in units of the Fermi wave
number $k_{F}$. Therefore it is convenient to measure real space distances
$\xi$ in units of $\lambda_{F}/2$, i.e, half the Fermi wave length. (In this
unit the wave length of the Friedel and RKKY oscillations is "$1$").

The density of the Wilson states $\psi_{\nu}\left(  \xi\right)  $ is given by%
\[
\rho_{\nu}^{0}\left(  \xi\right)  =\left\vert \psi_{\nu}\left(  \xi\right)
\right\vert ^{2}=2^{\nu+3}\frac{\sin^{2}\left(  \pi\xi\frac{1}{2^{\nu+2}%
}\right)  }{\pi\xi}d\xi
\]
(for $\nu<N/2$)$.$ The main contribution to the density $\rho_{\nu}^{0}$ of
the state $\psi_{\nu}$ lies roughly in the region $\left\vert \xi\right\vert
<$ $2^{\nu+2}$ (in units of $\lambda_{F}/2$). The different $\psi_{\nu}\left(
\xi\right)  $ have very different electron densities and vary by roughly a
factor of $2^{N/2}$ (which is generally larger than $10^{6}$). Therefore it is
useful to calculate the integrated electron density on a logarithmic scale.

First we discuss the magnetic state whose wave function is given in
equ.(\ref{y1}).

\subsection{The Magnetic State}

The magnetic state $\Psi_{MS}$ is the building block of the singlet state. Its
multi-electron state is built from four Slater states and shown in equ.
(\ref{y1}). The electron system has already a finite density without the
d-impurity. Therefore it is useful to calculate the change of the (integrated)
densities for spin-up and down conduction electrons due to the d-impurity.

In Fig.5 these (integrated) densities as well as their sum and difference
(total density and polarization) are plotted for the parameters $E_{d}=-0.5,$
$\left\vert V_{sd}^{0}\right\vert ^{2}=0.04,$ $U=1$ and $N=50$. With these
parameters the impurity has a well developed magnetic moment of $\mu
=0.93\mu_{B}$. The occupation of the different components is $A_{a,b}%
^{2}=0.0294,$ $A_{a,d}^{2}=0.0057,$ $A_{d,b}^{2}=0.9355$ and $A_{d,d}%
^{2}=0.0294.$ This means that $93.6$\% of the densities is due to the Slater
state $\Psi_{d,b}=d_{\uparrow}^{\dag}b_{0\downarrow}^{\dag}\prod_{i=1}%
^{n-1}a_{i\uparrow}^{\dag}\prod_{i=1}^{n-1}b_{i\downarrow}^{\dag}\Phi_{0}$.
The abscissa is the logarithm (using the basis 2) of $\xi=2x/\lambda_{F}$.

The region beyond $\xi=2^{20}$ corresponds to the rim or surface of the sample
and is discussed below. One recognizes that there is only a negligible
polarization of the electron gas in the vicinity of the impurity. The
important result of Fig.5 is that there is no polarization cloud around the
magnetic state of the impurity.
\begin{align*}
&
{\includegraphics[
height=3.2893in,
width=4.0872in
]%
{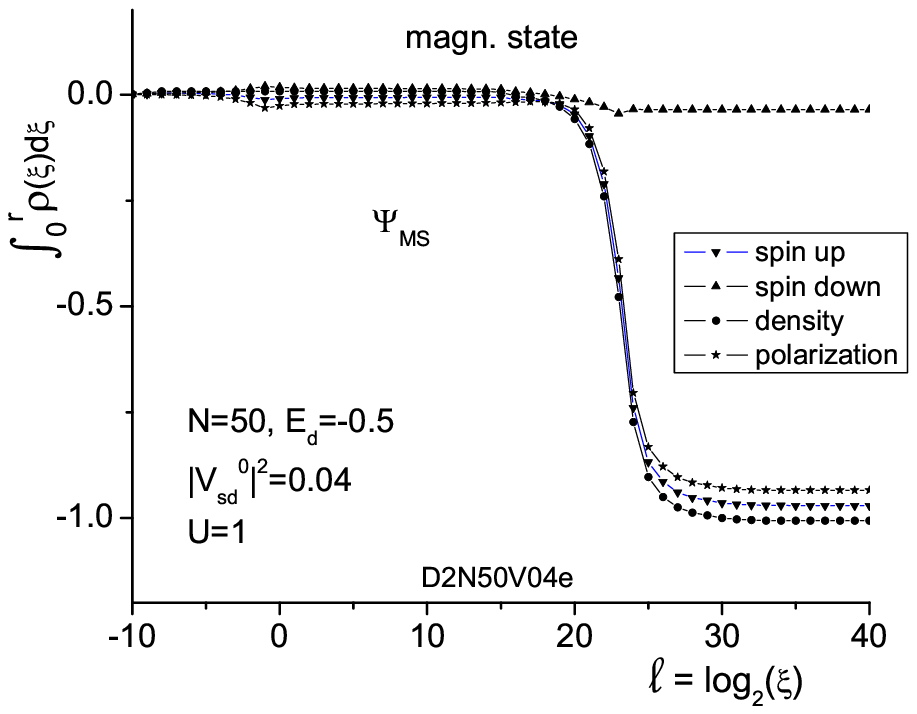}%
}%
\\
&
\begin{tabular}
[c]{l}%
Fig.5: The net integrated density $\int_{0}^{r}\rho\left(  \xi\right)  d\xi$
of the s-electron within a distance $r$\\
from the impurity for spin up and down, as well as total density and spin\\
polarization. The magnetic moment of the impurity is $0.93\mu_{B}$.
\end{tabular}
\end{align*}

If one looks at the magnetic state, in particular the dominant component
$\Psi_{d,b}=d_{\uparrow}^{\dag}b_{0\downarrow}^{\dag}\prod_{i=1}%
^{n-1}a_{i\uparrow}^{\dag}\prod_{i=1}^{n-1}b_{i\downarrow}^{\dag}\Phi_{0}$
then one realizes that the total magnetic moment of all the conduction
electrons is equal to $-1\mu_{B}$ (cancelling the moment of $+1\mu_{B}$ of the
d-electron which is not shown in Fig.5). So how can the state $\Psi_{MS}$ have
a finite magnetic moment. The answer is given by Fig.5. The moment $\mu_{B}$
of the s-electrons is pushed towards the surface of the sample which is at the
largest radius used for the Wilson states, i.e. $2^{N/2}$. This explains the
change of the integrated densities and polarization at $l\thickapprox22$ from
zero to $-1$. If one increases the number $N$ of Wilson states by $\Delta N$
then the transition is shifted by $\Delta l=\Delta N/2$.

\subsection{The Kondo cloud}

One of the most controversial aspects of the Kondo ground state is the
so-called Kondo cloud within the radius $r_{K}$ where $r_{K}$ is called the
Kondo length
\begin{equation}
r_{K}=\frac{\hbar v_{F}}{\varepsilon_{K}}=\frac{d\varepsilon/dk}%
{\varepsilon_{K}}%
\end{equation}
($\varepsilon_{K}$= Kondo energy, $v_{F}$ = Fermi velocity of the
s-electrons). For a linear dispersion relation this yields in reduced units
$\xi_{K}=1/\left(  \pi\varepsilon_{K}\right)  $ where in this relation the
Kondo energy is given in units of the half-band width.

The idea is to divide the ground state $\Psi_{K}$ of a Kondo impurity into two
parts with opposite d-spins. The proponents of the Kondo cloud argue that in
each component there is an s-electon cloud within the Kondo sphere which
compensates the d-spin. An important assumption of the Kondo-cloud proponents
is that, above the Kondo temperature, the bond is broken and this screening
cloud evaporates from the Kondo sphere.

In the 1970's Slichter and co-workers \cite{S84} investigated Cu samples with
dilute Fe-Kondo impurities by means of nuclear magnetic resonance. They did
not detect any Kondo cloud. In a number of recent theoretical papers
\cite{A81}, \cite{A80}, \cite{A82}, \cite{S83} the argument is made that the
old NMR experiments could not possibly have detected the screening electron
because of the large volume of the Kondo sphere yielding a polarization of
less than $10^{-8}$ electron spins per host atom.

The FAIR solution of the singlet ground state is well suited to determine the
electron density and polarization in real space \cite{B177}. To the knowledge
of the author this is the first detailed calculation of the Kondo cloud.

In the following analysis of the singlet state $\Psi_{SS}$ the same parameters
are used as for the magnetic state in Fig.5: $E_{d}=-0.5,$ $\left\vert
V_{sd}^{0}\right\vert ^{2}=0.04$, $U=1$. This yields the following squared
amplitudes: $\overline{A_{a,b}^{2}}=0.0146,$ $\overline{A_{a,d}^{2}}=0.0028,$
$\overline{A_{db}^{2}}=0.4629$ and $\overline{A_{dd}^{2}}=0.0146$. (The
$\overline{\overline{A_{x,y}}}$ amplitudes are identical). These occupations
are very close to half the values of the magnetic state ($A_{s,s}^{2}=0.0294,$
$A_{s,d}^{2}=0.0057,$ $A_{d,s}^{2}=0.9355$ and $A_{d,d}^{2}=0.0294)$. This
means that $\Psi_{SS}$ is given in good approximation as $\Psi_{SS}%
\thickapprox\left(  1/\sqrt{2}\right)  \left[  \overline{\Psi_{MS}\left(
\uparrow\right)  }+\overline{\overline{\Psi_{MS}\left(  \downarrow\right)  }%
}\right]  $. The two magnetic states with d-spin up and down are robust
building blocks of the singlet state. (However, there are subtle changes in
the FAIR states which will be discussed below). Therefore the spin
polarization of one of the magnetic components, for example of $\overline
{\Psi_{MS}\left(  \uparrow\right)  }$, would be of interest.

In Fig.6 the integrated densities of spin up and down electrons, their sum and
difference (the polarization) are plotted versus the distance from the
magnetic impurity (on a logarithmic scale). One recognizes that now one has
considerable contributions to the integrated net densities of both spins. The
polarization of the two contributions is no longer zero but reaches a value of
$-0.46$ at a distance of $r\thickapprox2^{1\text{1.6}}$. Since the magnetic
state $\overline{\Psi_{MS}\left(  \uparrow\right)  }$ with net d-spin up has
only a weight of about $1/2$ it contributes an effective $d_{\uparrow}^{\dag}%
$-moment of $0.93/2\thickapprox0.46$. Therefore this d-spin is well
compensated by the polarization of the s-electron background.

The difference with the pure magnetic state is particularly striking. We
observe a screening polarization cloud of s-electrons about the impurity
within the range of $\xi\thickapprox2^{1\text{1.6}}$ $\ $or $r=\allowbreak
3.1\times10^{3}\left(  \lambda_{F}/2\right)  .$This is about the Kondo length
$r_{K}$.
\begin{align*}
&
{\includegraphics[
height=3.4147in,
width=4.1735in
]%
{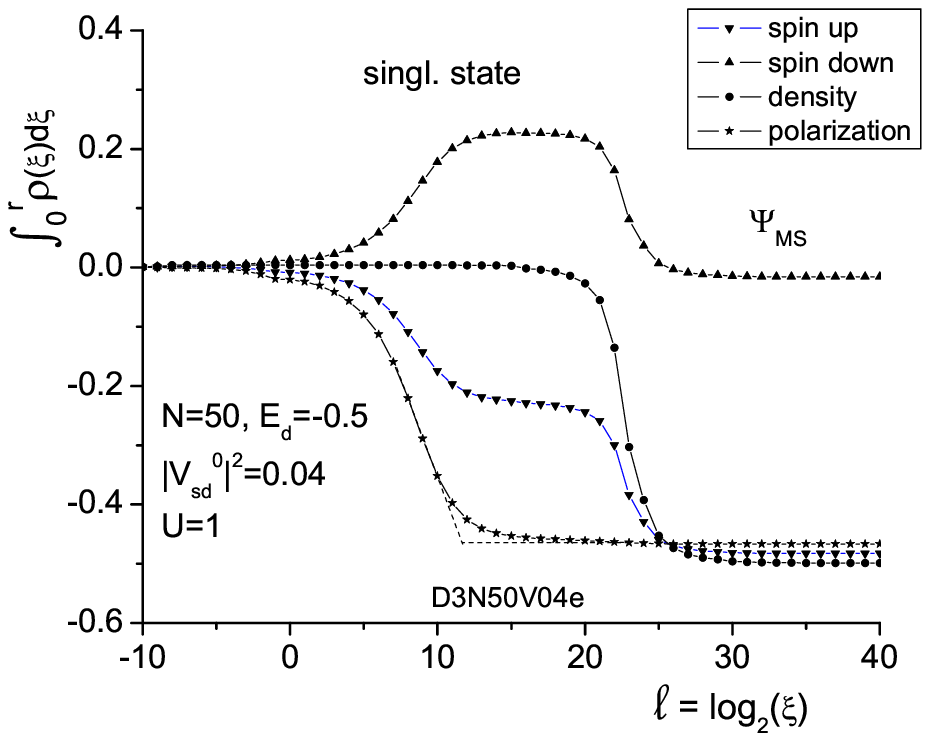}%
}%
\\
&
\begin{tabular}
[c]{l}%
Fig.6: The net integrated density $\int_{0}^{r}\rho\left(  x\right)  dx$
within a distance $\xi=2^{l}$\\
from the d-spin up component of the impurity$.$ Shown are the spin up, spin\\
down components as well as the total density and the polarization. The
d$_{\uparrow}$-spin\\
of $0.93/2$ is screened by 0.46 s-electrons within the range of $\xi
\thickapprox2^{11.6}$\\
(or $r\thickapprox3\times10^{3}\lambda_{F}/2$ ).
\end{tabular}
\end{align*}%
\[
\]

The polarization cloud for the ground state of the Kondo impurity
(\ref{Psi_K}) is in principle identical with the results for the
Friedel-Anderson impurity and is discussed in detail in ref. \cite{B177}.

\subsection{Friedel Oscillation}

Recently Affleck, Borda and Saleur (ABS) \cite{A83} showed that the Friedel
oscillations due to a Kondo impurity are essentially suppressed within a
distance of the order of the Kondo length $r_{K}$. They supported their theory
by numerical calculations using NRG. Fig.7a shows the universal behavior of
their numerical results for many different interaction strengths. Plotted is a
function $F\left(  \xi/\xi_{K}\right)  $ (The actual amplitude is proportional
to $\left[  1-F\right]  \xi^{-D}$ where $D$ is the dimension of the system).
For $F=1$ the Friedel oscillation is canceled while for $F=-1$ its amplitude
is doubled). The author could not resist the temptation to evaluate the
Friedel oscillations with the FAIR method \cite{B178}. Fig.7b shows the FAIR
results of the amplitude $\left(  1-F\right)  $ of the Friedel oscillation for
two different interaction strengths. This universal curve is shown in Fig.7a
as the full blue curve. It agrees well with the numerical results by ABS. (The
exact form of $F\left(  \xi/\xi_{K}\right)  $ is not known explicitly).%

\[%
\begin{array}
[c]{c}%
{\includegraphics[
height=2.1262in,
width=2.7381in
]%
{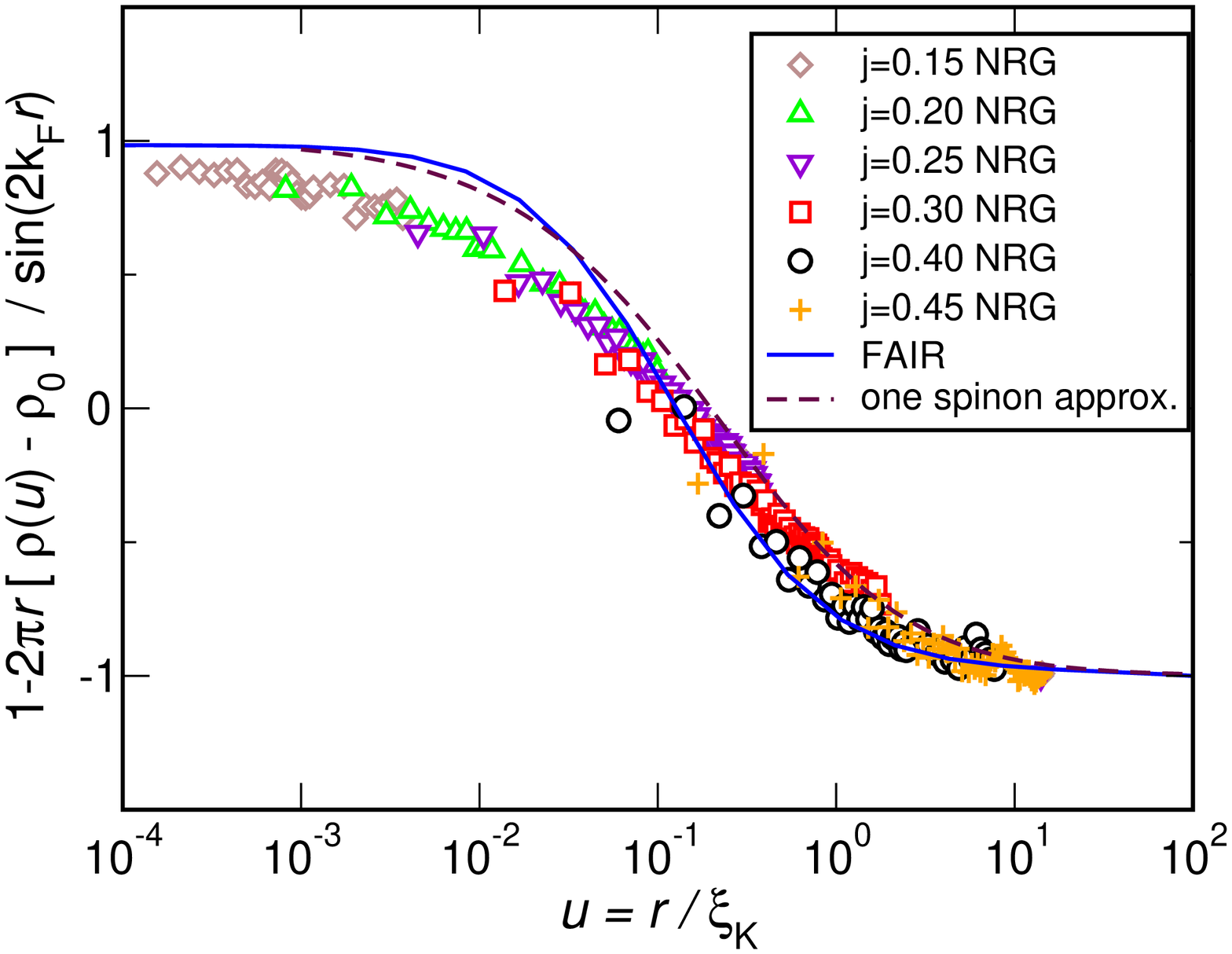}%
}%
\\%
\begin{tabular}
[c]{l}%
Fig.7a: Reduction $F\left(  \xi/\xi_{K}\right)  $ of the\\
Friedel oscillation is plotted\\
versus $\xi/\xi_{K}$. The amplitude\\
is proportional to $\left[  1-F\left(  \xi/\xi_{K}\right)  \right]  .$\\
Calculated by ABS.
\end{tabular}
\end{array}%
\begin{array}
[c]{c}%
{\includegraphics[
height=2.5687in,
width=2.8908in
]%
{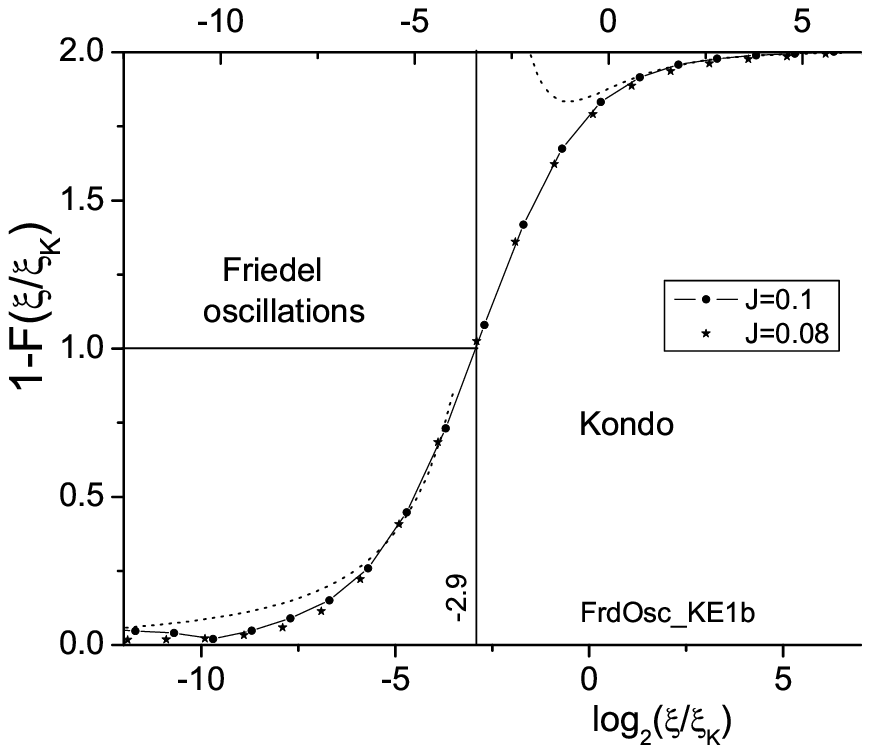}%
}%
\\%
\begin{tabular}
[c]{l}%
Fig.7b: The amplitude $\left[  1-F\left(  \xi/\xi_{K}\right)  \right]  $ of\\
the Friedel oscillation for two different\\
Kondo energies (from FAIR).\\
\end{tabular}
\end{array}
\]

\[
\]

\section{The FAIR States}

At the heart of the FAIR approach are the FAIR states $a_{0}^{\dag}$ and
$b_{0}^{\dag}$. Therefore a detailed discussion of these states is
appropriate. The FAIR states are expressed in terms of the Wilson states. The
latter are described in appendix A and a basic knowledge is required to follow
some of the arguments of this paragraph.

A FAIR state is given as $a_{0}^{\dag}=%
{\textstyle\sum_{\nu}}
\alpha_{0}^{\nu}c_{\nu}^{\dag}$ where the states $c_{\nu}^{\dag}$ are Wilson
states. Now each Wilson state $c_{\nu}^{\dag}$ represents all the original
band states $\varphi_{k}^{\dag}$ within the energy cell $\mathfrak{C}_{\nu}$
with an energy width $\Delta_{\nu}$ where $\Delta_{\nu}=\left(  \zeta_{\nu
+1}-\zeta_{\nu}\right)  $. The composition of $\ $the Wilson states $c_{\nu
}^{\dag}$ is $c_{\nu}^{\dag}$=$Z_{\nu}^{-1/2}%
{\textstyle\sum_{k}}
\varphi_{k}^{\dag}$. This yields for the FAIR state the composition%
\[
a_{0}^{\dag}=%
{\textstyle\sum_{\nu}}
{\textstyle\sum_{k}}
\frac{\alpha_{0}^{\nu}}{\sqrt{Z_{\nu}}}\varphi_{k}^{\dag}%
\]
This means that the FAIR state $a_{0}^{\dag}$ consists of the original s-band
states $\varphi_{k}^{\dag}$ which have the amplitude of $\alpha_{0}^{\nu
}/\sqrt{Z_{\nu}}$ in the energy cell $\mathfrak{C}_{\nu}$ or the occupation
$\left\vert \alpha_{0}^{\nu}\right\vert ^{2}/Z_{\nu}$. Now we can express
\[
\frac{\left\vert \alpha_{0}^{\nu}\right\vert ^{2}}{Z_{\nu}}=\frac{1}{Z}%
\frac{Z}{Z_{\nu}}\left\vert \alpha_{0}^{\nu}\right\vert ^{2}=\frac{2}{Z}%
\frac{\left\vert \alpha_{0}^{\nu}\right\vert ^{2}}{\Delta_{\nu}}%
\]
where $Z$ is the total number of $\varphi_{k}^{\dag}$ states in the conduction
electron band (for one spin) and $Z/Z_{\nu}=2/\Delta_{\nu}$

Therefore the expression $p_{\nu}=$ $\left\vert \alpha_{0}^{\nu}\right\vert
^{2}/\Delta_{\nu}$ represents (besides the factor $Z/2$) the composition of
the FAIR state $a_{0}^{\dag}$ in terms of the original band state $\varphi
_{k}^{\dag}$ in the energy cell $\mathfrak{C}_{\nu}.$

If one plots $p_{\nu}$ as a function of energy then one finds a step function
because of the finite energy width of the cells $\mathfrak{C}_{\nu}$ of the
Wilson states. If one repeatedly sub-divides the energy cells (doubling the
number of Wilson states) then a smooth function $p\left(  \zeta\right)  $
emerges. This yields the FAIR state $a_{0}^{\dag}$ ($b_{0}^{\dag}$) in a
quasi-continuous energy band. A rather good approximation of $p\left(
\zeta\right)  $ can be obtained by interpolation.

A comparison of Fig.5 and Fig.6 for the polarization of the magnetic state
$\Psi_{MS}\left(  \uparrow\right)  $ and the magnetic component $\overline
{\Psi_{MS}\left(  \uparrow\right)  }$ of the singlet state $\Psi_{SS}$ shows a
remarkable difference in the polarization about the impurity although the
structure of the two states is identical. This is particularly surprising
since the coefficients $\overline{A_{\alpha,\beta}}=\overline{\overline
{A_{\alpha,\beta}}}$ in the singlet state are roughly $1/\sqrt{2}$ of the
coefficients $A_{\alpha,\beta}$ of the magnetic state. However, in the singlet
state one has a finite coupling between $\overline{\Psi_{MS}\left(
\uparrow\right)  }$ and $\overline{\overline{\Psi_{MS}\left(  \downarrow
\right)  }}$. This shifts the composition of the FAIR states $a_{0}^{\dag}$
and $b_{0}^{\dag}$ towards small energies. The difference is that the FAIR
states in the two states have a very different composition. To demonstrate
this difference in the compositions $p_{\nu}=$ $\left\vert \alpha_{0}^{\nu
}\right\vert ^{2}/\Delta_{\nu}$ of $a_{0}^{\dag}$ (and $b_{0}^{\dag})$ are
plotted in Fig.8 as a function of $\nu$ for the two different states.%

\begin{align*}
&
\begin{array}
[c]{cc}%
{\includegraphics[
height=2.1162in,
width=2.6866in
]%
{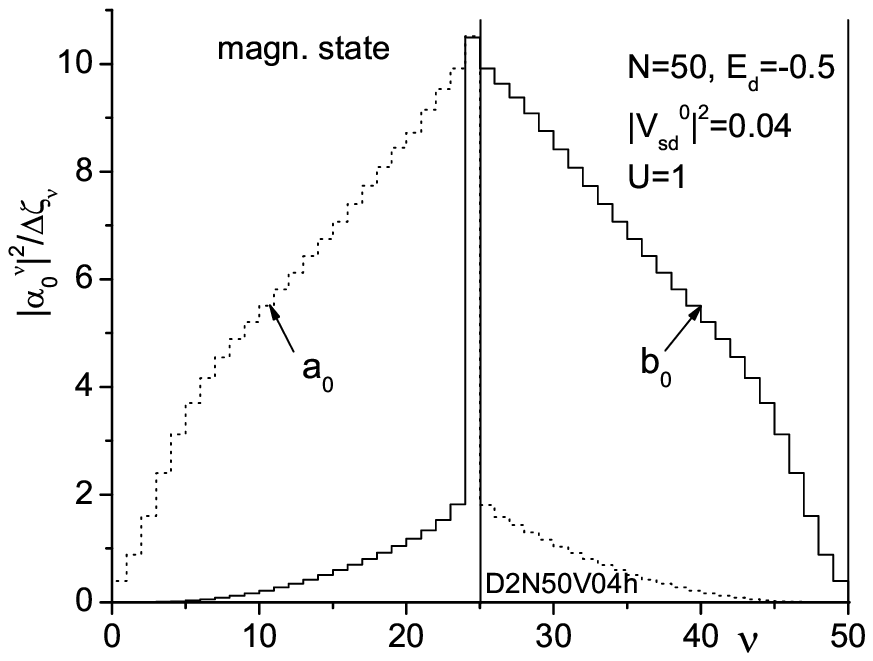}%
}%
&
{\includegraphics[
height=2.122in,
width=2.7978in
]%
{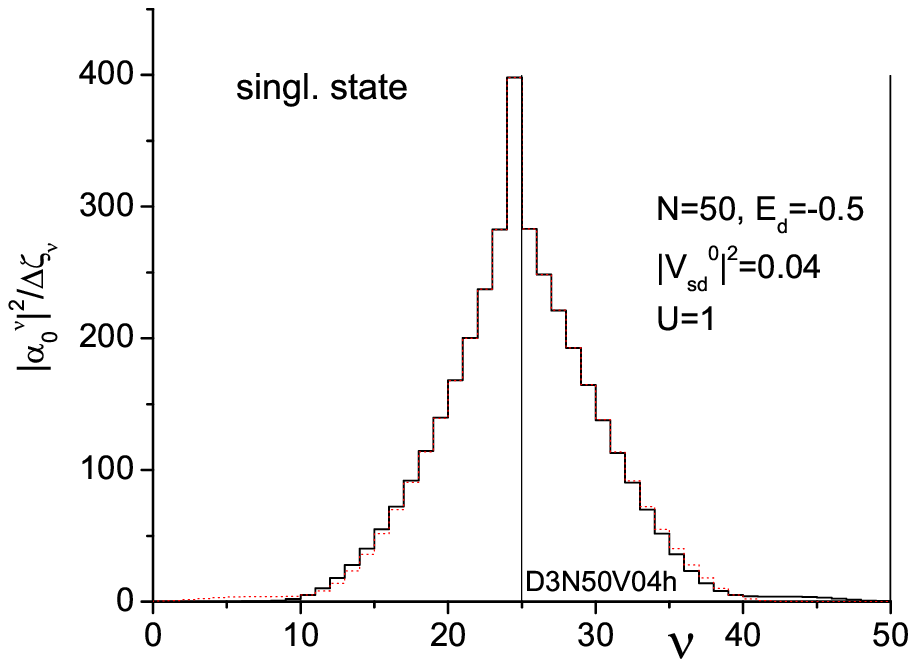}%
}%
\end{array}
\\
&
\begin{tabular}
[c]{l}%
Fig.8a: The density distribution of the $p_{\nu}=$ $\left\vert \alpha_{0\pm
}^{\nu}\right\vert ^{2}/\Delta_{\nu}$ for the magnetic state $\Psi_{MS}$\\
as a function of as a function of $\nu.$\\
Fig.8b: The density distribution of the $p_{\nu}=$ $\left\vert \alpha_{0\pm
}^{\nu}\right\vert ^{2}/\Delta_{\nu}$ for the singlet ground\\
state $\Psi_{SS}$ as a function of the cell number $\nu.$ Note the difference
in scale.
\end{tabular}
\end{align*}

It would be more natural to plot $p_{\nu}$ as a function of the energy
$p\left(  \zeta\right)  $. But for $\nu$ close to $N/2$ the width of the
energy cells $\mathfrak{C}_{\nu}$ is less than $10^{-6}$ and any dependence of
$p\left(  \zeta\right)  $ on the energy cannot be resolved on a linear scale.
The probability $p_{\nu}$ increases close to the Fermi energy.

In Fig.8a for the magnetic state the compositions of $a_{0}^{\dag}$ and
$b_{0}^{\dag}$ resemble mirror images. The function $p_{\nu}$ has a maximum at
small energies of about $10$.

In Fig.8b the corresponding plot is shown for the singlet state. The weight
$p_{\nu}=\left\vert \alpha_{0}^{\nu}\right\vert ^{2}/\Delta_{\nu}$ close to
the Fermi level is very different for the singlet state and the magnetic
state. One observes in the singlet state a maximum of about 400 and the
weights $p_{\nu}$ in $a_{0}^{\dag}$ and $b_{0}^{\dag}$ are essentially
identical and not mirror images. The magnetic component of the singlet state
is in a subtle way different from the magnetic state.

It should be emphasized again that the two states $a_{0}^{\dag}$ and
$b_{0}^{\dag}$ contain the whole information about the many electron states
$\Psi_{MS}$ or $\Psi_{SS}$. When $a_{0}^{\dag}$ and $b_{0}^{\dag}$ are known
the whole bases $\left\{  a_{i}^{\dag}\right\}  $and $\left\{  b_{i}^{\dag
}\right\}  $ and all the coefficients $A_{\alpha,\beta}$ or $\overline
{A_{\alpha,\beta}},\overline{\overline{A_{\alpha,\beta}}}$ can be reconstructed.

One important advantage of the FAIR method is that one can modify the size of
the energy cells $\mathfrak{C}_{\nu}$ after one has completed the numerical
calculation of the FAIR states and the ground state. It requires relatively
little effort to sub-divide the Wilson states. This is an important advantage
over NRG which can't change the energy width of the Wilson energy cells
$\mathfrak{C}_{\nu}.$ This procedure is discussed in appendix C.

\appendix{}

\section{Wilson's s-electron basis}

Wilson \cite{W18} in his Kondo paper considered an s-band with a constant
density of states and the Fermi energy in the center of the band. By measuring
the energy from the Fermi level and dividing all energies by the Fermi energy
Wilson obtained a band ranging from $-1$ to $+1.$ To treat the electrons close
to the Fermi level at $\zeta=0$ as accurately as possible he divided the
energy interval $\left(  -1:0\right)  $ at energies of $-1/\Lambda
,-1/\Lambda^{2},-1/\Lambda^{3},..$ i.e. $\zeta_{\nu}=-1/\Lambda^{\nu}.$ This
yields energy cells $\mathfrak{C}_{\nu}$ with the range $\left\{
-1/\Lambda^{\nu-1}:-1/\Lambda^{\nu}\right\}  $ and the width $\Delta_{\nu}$
$=\zeta_{\nu}-\zeta_{\nu-1}$ $=1/\Lambda^{\nu}$. Generally the value
$\Lambda=2$ is chosen. (There are equivalent intervals for positive $\zeta
$-values where $\nu$ is replaced by $\left(  N-\nu\right)  $ but we discuss
here only the negative energies). The new Wilson states $c_{\nu}^{\ast}$ are a
superposition of all states in the energy interval $\left(  \zeta_{\nu
-1},\zeta_{\nu}\right)  $ and have an (averaged) energy $\left(  \zeta_{\nu
}+\zeta_{\nu-1}\right)  /2=\allowbreak\left(  -\dfrac{3}{2}\right)  \dfrac
{1}{2^{\nu}}$, i.e. $-\frac{3}{4},-\frac{3}{8},-\frac{3}{16},..,-\frac
{3}{2\cdot2^{N/2}},-\frac{1}{2\cdot2^{N/2}}.$ (I count the energy cells and
the Wilson states from $\nu=1$ to $N$). This spectrum continues symmetrically
for positive energies. The essential advantage of the Wilson basis is that it
has an arbitrarily fine energy spacing at the Fermi energy.

Wilson rearranged the original quasi-continuous electron states $\varphi
_{k}^{\dag}$ in such a way that only one state within each cell $\mathfrak{C}%
_{\nu}\ $had a finite interaction with the impurity. Assuming that the
interaction of the original electron states $\varphi_{k}^{\dag}$ with the
impurity is $k$-independent this interacting state in $\mathfrak{C}_{\nu}$ had
the form%
\[
c_{\nu}^{\dag}=%
{\textstyle\sum_{\mathfrak{C}_{\nu}}}
\varphi_{k}^{\dag}/\sqrt{Z_{\nu}}%
\]
where $Z_{\nu}$ is the total number of states $\varphi_{k}^{\dag}$ in the cell
$\mathfrak{C}_{\nu}$ ($Z_{\nu}=Z\left(  \zeta_{\nu}-\zeta_{\nu-1}\right)  /2,$
$Z$ is the total number of states in the band). There are $\left(  Z_{\nu
}-1\right)  $ additional linear combinations of the states $\varphi_{k}^{\dag
}$ in the cell $\mathfrak{C}_{\nu}$ but they have zero interaction with the
impurity and were ignored by Wilson, as they are within this work.

\section{Construction of the Basis $a_{0}^{\dag}$, $a_{i}^{\dag}$}

For the construction of the state $a_{0}^{\dag}$ and the rest of basis
$a_{i}^{\dag}$ one starts with the s-band electrons $\left\{  c_{\nu}^{\dag
}\right\}  $ which consist of $N$ states (for example Wilson's states). The
$d^{\dag}$-state is ignored for the moment. \newline

\begin{itemize}
\item In step (1) one forms a normalized state $a_{0}^{\dag}$ out of the
s-states with:
\end{itemize}

\begin{equation}
a_{0}^{\dag}=\sum_{\nu=1}^{N}\alpha_{0}^{\nu}c_{\nu}^{\dag}%
\end{equation}
The coefficients $\alpha_{0}^{\nu}$ can be arbitrary at first. One reasonable
choice is $\alpha_{0}^{\nu}=1/\sqrt{N}$

\begin{itemize}
\item In step (2) $\left(  N-1\right)  $ new basis states $\overline{a}%
_{i}^{\dag}$ $\left(  1\leq i\leq N-1\right)  $ are formed which are
normalized and orthogonal to each other and to $a_{0}^{\dag}$.

\item In step (3) the s-band Hamiltonian $H_{0}$ is constructed in this new
basis. One puts the state $a_{0}^{\dag}$ at the top so that its matrix
elements are $H_{0i}$ and $H_{i0}$.

\item In step (4) the $\left(  N-1\right)  $-sub Hamiltonian which does not
contain the state $a_{0}^{\dag}$ is diagonalized. This transforms the rest of
the basis $\left\{  \overline{a}_{i}^{\dag}\right\}  $ into a new basis
$\left\{  a_{0}^{\dag},a_{i}^{\dag}\right\}  $ (but keeps the state
$a_{0}^{\dag}$ unchanged). The resulting Hamilton matrix for the s-band then
has the form%
\begin{equation}
H_{0}=\left(
\begin{array}
[c]{ccccc}%
E(0) & V_{fr}(1) & V_{fr}(2) & ... & V_{fr}(N-1)\\
V_{fr}(1) & E(1) & 0 & ... & 0\\
V_{fr}(2) & 0 & E(2) & ... & 0\\
.. & ... & ... & ... & ...\\
V_{fr}(N-1) & 0 & 0 & ... & E(N-1)
\end{array}
\right)  \label{hmat}%
\end{equation}
The creation operators of the new basis are given by the set $\left\{
a_{0}^{\dag},a_{i}^{\dag}\right\}  ,$ ($0<i\leq N-1)$. The $a_{i}^{\dag}$ can
be expressed in terms of the s-states; $a_{i}^{\dag}=\sum_{\nu=1}^{N}%
\alpha_{i}^{\nu}c_{\nu}^{\dag}$. The state $a_{0}^{\dag}$ uniquely determines
the other states $a_{i}^{\dag}$. The state $a_{0}^{\dag}$ is coupled through
the matrix elements $V_{fr}\left(  i\right)  $ to the states $a_{i}^{\dag}$,
which makes the state $a_{0}^{\dag}$ an artificial Friedel resonance. The
matrix elements $E\left(  i\right)  $ and V$_{fr}\left(  i\right)  $ are given
as%
\begin{align*}
E(i)  &  =\sum_{\nu}\alpha_{i}^{\nu}\varepsilon_{\nu}\alpha_{i}^{\nu}\\
V_{fr}\left(  i\right)   &  =%
{\textstyle\sum_{\nu}}
\alpha_{0}^{\nu}\varepsilon_{\nu}\alpha_{i}^{\nu}%
\end{align*}

\item In the final step (5) the state $a_{0}^{\dag\text{ }}$is rotated in the
$N$-dimensional Hilbert space. In each cycle the state $a_{0}^{\dag}$ is
rotated in the $\left(  a_{0}^{\dag}\text{,}a_{i_{0}}^{\dag}\right)  $ plane
by an angle $\theta_{i_{0}}$ for $1\leq i_{0}\leq N-1$. Each rotation by
$\theta_{i_{0}}$ yields a new $\overline{a_{0}}^{\dag}$
\[
\overline{a_{0}}^{\dag}=a_{0}^{\dag}\cos\theta_{i_{0}}+a_{i_{0}}^{\dag}%
\sin\theta_{i_{0}}%
\]

The rotation leaves the whole basis $\left\{  a_{0}^{\dag},a_{i}^{\dag
}\right\}  $ orthonormal. Step (4), the diagonalization of the $\left(
N-1\right)  $-sub Hamiltonian, is now much quicker because the $\left(
N-1\right)  $-sub-Hamiltonian is already diagonal with the exception of the
$i_{0}$- row and the $i_{0}$-column . For each rotation plane $\left(
a_{0}^{\dag}\text{,}a_{i_{0}}^{\dag}\right)  $ the optimal $a_{0}^{\dag}$ with
the lowest energy expectation value is determined. This cycle is repeated
until one reaches the absolute minimum of the energy expectation value. In the
example of the Friedel resonance Hamiltonian this energy agrees numerically
with an accuracy of $10^{-15}$ with the exact ground-state energy of the
Friedel Hamiltonian \cite{B91}. For the Kondo impurity the procedure is
stopped when the expectation value changes by less than $10^{-10}$ during a
full cycle.
\end{itemize}

\section{Changing the Wilson Basis}

In NRG one usually constructs the Wilson states with $\Lambda=2$. This means
that one uses energy cells whose width reduced by a factor of two. NRG is in
principle exact for $\Lambda\thickapprox1$ (together with the requirement that
one can diagonalize matrices of gigantic sizes). In the FAIR method we also
begin the calculation with $\Lambda=2$. When the FAIR state $a_{0}^{\dag}$ is
obtained for $\Lambda=2$ in the basis $\left\{  c_{\nu}^{\dag}\right\}  $ then
it is also approximately known in the original basis $\left\{  \varphi
_{k}^{\dag}\right\}  $ (with $10^{23}$ states)
\[
a_{0}^{\dag}=%
{\textstyle\sum_{\nu=1}^{N}}
\alpha_{0}^{\nu}c_{\nu}^{\dag}=%
{\textstyle\sum_{\nu=1}^{N}}
\alpha_{0}^{\nu}%
{\textstyle\sum_{\mathfrak{C}_{\nu}}}
\varphi_{k}^{\dag}/\sqrt{Z_{\nu}}%
\]
Now one can choose a smaller $\Lambda$, for example $\Lambda=\sqrt{2}$ and
interpolate with good accuracy the FAIR state for the smaller value of
$\Lambda$ \cite{B178}. The optimization of the resulting FAIR state requires
now a relatively short additional numerical iteration. For the calculation of
the Friedel oscillation a value of $\Lambda=\sqrt[4]{2}=\allowbreak
1.\,\allowbreak19$ was used.

\section{Geometrical derivation of the Friedel ground state}

If the conduction electrons are described by a basis of $N$ states then
together with the d-state this yields an $\left(  N+1\right)  $-dimensional
Hilbert space $\mathfrak{H}_{N+1}$. The Friedel Hamiltonian is a single
particle Hamiltonian and possesses in our case $\left(  N+1\right)  $
orthonormal eigenstates $b_{j}^{\dag}$ which are compositions of the $N$
states $c_{\nu}^{\dag}$ and the one d state $d^{\dag}$. The $n$-electron
ground state is then the product of the $n$ creation operator $b_{j}^{\dag}$
with the lowest energy (applied to the vacuum state $\Phi_{0}$). These $n$
states define the $n$-dimensional occupied sub-Hilbert space $\mathfrak{H}%
_{n}$. The remaining $\left(  N+1-n\right)  $ eigenstates form the
complementary unoccupied sub-Hilbert space $\mathfrak{H}_{N+1-n}$. In the
following we treat the creation operators as unit vectors within the Hilbert space.

Now the vector $\mathbf{d}$ of the d state lies partially in the occupied and
the unoccupied sub-Hilbert space. It has a projection $\mathbf{d}_{1}^{\prime
}$ in the occupied sub-Hilbert space $\mathfrak{H}_{n}$ and a projection
$\mathbf{d}_{2}^{\prime}$ in the unoccupied sub-Hilbert space $\mathfrak{H}%
_{N+1-n}$ (so that $\mathbf{d}=\mathbf{d}_{1}^{\prime}+\mathbf{d}_{2}^{\prime
}$). The lengths of the vectors $\mathbf{d}_{1}^{\prime}$ and $\mathbf{d}%
_{2}^{\prime}$ are less than one. So we normalize them to $\mathbf{d}_{1}$ and
$\mathbf{d}_{2}$ with $\left\vert \mathbf{d}_{i}\right\vert =1$. These two
vectors are orthogonal (they lie in different sub-Hilbert spaces) and form
therefore a two-dimensional space. The vector $\mathbf{d}$ lies within this
plane because
\[
\mathbf{d=d}_{1}^{\prime}+\mathbf{d}_{2}^{\prime}=\alpha\mathbf{d}_{1}%
+\beta\mathbf{d}_{2}%
\]
The vector perpendicular to $\mathbf{d}$ in this plane is the FAIR state
$\mathbf{a}_{0}$ with the composition
\[
\mathbf{a}_{0}=\beta\mathbf{d}_{1}-\alpha\mathbf{d}_{2}%
\]
Then the vector $\mathbf{d}_{1}$ has the form
\[
\mathbf{d}_{1}=\beta\mathbf{a}_{0}+\alpha\mathbf{d}%
\]

The vector $\mathbf{d}_{1}$ can be used as a basis vector of the $\left(
N+1\right)  $ Hilbert space $\mathfrak{H}_{N+1}$. It lies completely within
the occupied sub-Hilbert space. Now we divide the occupied sub-Hilbert space
$\mathfrak{H}_{n}$ into the one-dimensional space $\mathbf{d}_{1}$ and an
$\left(  n-1\right)  $-dimensional subspace $\mathfrak{S}_{n-1}$ which is
orthogonal to $\mathbf{d}_{1}$. This subspace $\mathfrak{S}_{n-1}$ is also
orthogonal to the d state vector $\mathbf{d}$ and is therefore built only of
$\mathbf{c}_{\nu}$ vectors. It can be decomposed into $\left(  n-1\right)  $
orthonormal basis vectors $\overline{\mathbf{a}}_{i}$.

Returning to the physics, the ground state can be expressed as
\[
\Psi_{F}=\left(  \beta a_{0}^{\dag}+\alpha^{d}\dag\right)
{\textstyle\prod\limits_{i=1}^{n-1}}
\overline{a}_{i}^{\dag}\Phi_{0}%
\]

Similarly the sub-Hilbert space $\mathfrak{H}_{N+1-n}$ can be divided into
$\mathbf{d}_{2}$ and a sub-space $\mathfrak{S}_{N-n}$ orthogonal to
$\mathbf{d}_{2}$ which is therefore also orthogonal to $\mathbf{d}$.
$\mathfrak{S}_{N-n}$ can be expressed in terms of $\left(  N-n\right)  $
orthonormal basis vectors $\overline{\mathbf{a}}_{i}$ (which consists only of
vectors $\mathbf{c}_{\nu}$).

The creation operators $\overline{a}_{i}$ are not yet uniquely determined.
That is done by diagonalizing the Hamiltonian $H_{0}$ in $\mathfrak{S}_{n-1}$
and $\mathfrak{S}_{N-n}$. This yields the new basis $\left\{  a_{i}^{\dag
},1\leq i<N-1\right\}  $. It is straight forward to show that the matrix
elements $\left\langle a_{i}^{\dag}\left\vert H_{0}\right\vert a_{i^{\prime}%
}^{\dag}\right\rangle $ for $a_{i}^{\dag}\in\mathfrak{S}_{n-1}$ and
$a_{i^{\prime}}^{\dag}\in\mathfrak{S}_{N-n}$ vanish as well. (We know the
matrix elements of $H_{F}$ between any state in $\mathfrak{S}_{N-n}$ and any
state in $\mathfrak{S}_{n-1}$ vanishes because the two sub-Hilbert spaces are
built from a different sub-set of eigenstates of $H_{F}$. Therefore
$\left\langle a_{i}^{\dag}\left\vert H_{F}\right\vert a_{i^{\prime}}^{\dag
}\right\rangle =0$ if $a_{i}^{\dag}\in\mathfrak{S}_{n-1}$ and $a_{i^{\prime}%
}^{\dag}\in\mathfrak{S}_{N-n}$. Since $\mathfrak{S}_{n-1}$\ and $\mathfrak{S}%
_{N-n}$ are orthogonal to $d^{\dag}$ the d component of the Hamiltonian
$H_{F}$ vanishes anyhow and the remaining part $\left\langle a_{i}^{\dag
}\left\vert H_{0}\right\vert a_{i^{\prime}}^{\dag}\right\rangle =0$ vanishes
for all pairs of $i$ and $i^{\prime}$.)

\end{document}